\documentclass[]{aa}
\usepackage[utf8]{inputenc}
\usepackage{graphicx}
\usepackage{txfonts}
\begin{document}

   \title{A jet-gas interaction beyond the host galaxy: Detection of a neutral hydrogen outflow at cosmic noon}

   \author{Renzhi Su \inst{1}, 
          Stephen J. Curran\inst{2}, 
          James R. Allison\inst{3}, 
          Marcin Glowacki\inst{4,5,6}, 
          Minfeng Gu\inst{1}, 
          Vanessa Moss\inst{7},
          Yongjun Chen\inst{1,8},
          Zhongzu Wu\inst{9},          
          \and
          Zheng Zheng\inst{10}          
          }

   \institute{Shanghai Astronomical Observatory, Chinese Academy of Sciences, 80 Nandan Road, Shanghai 200030, China\\
              \email{rzsu.astro@gmail.com; gumf@shao.ac.cn}
              \and
             School of Chemical and Physical Sciences, Victoria University of Wellington, PO Box 600, Wellington 6140, New Zealand
            \and
            First Light Fusion Ltd., Unit 9/10 Oxford Pioneer Park, Mead Road, Yarnton, Kidlington OX5 1QU, UK
            \and  
            Institute for Astronomy, University of Edinburgh, Royal Observatory, Edinburgh, EH9 3HJ, UK
            \and
            Inter-University Institute for Data Intensive Astronomy, Department of Astronomy, University of Cape Town, Cape Town, South Africa
            \and
            International Centre for Radio Astronomy Research (ICRAR), Curtin University, Bentley, WA, Australia
            \and
            ATNF, CSIRO Space and Astronomy, PO Box 76, Epping, NSW 1710, Australia
            \and
            State Key Laboratory of Radio Astronomy and Technology, A20 Datun Road, Chaoyang District, Beijing, 100101, P. R. China
            \and 
            College of Physics, Guizhou University, 550025 Guiyang, PR China
            \and
            National Astronomical Observatories, Chinese Academy of Sciences, 20A Datun Road, Beijing 100101, China
             }

   \date{Received XX XX, 202X; accepted XX XX, 202X}

 
\abstract
   {We present upgraded Giant Metrewave Radio Telescope (uGMRT) observations of 0731+438, an \mbox{FR II} radio galaxy at a redshift of 2.429 with two lobes separated by 82 kpc. A blueshifted, faint, and broad \mbox{H{\sc i}} 21 cm absorption line with a velocity full width at half maximum of $\sim 600\,\rm km\,s^{-1}$ is detected against the southern radio lobe that is 47 kpc from radio core, indicating a neutral hydrogen outflow associated with jet-gas interaction beyond the host galaxy. The outflow has a mass outflow rate of $\sim\,0.4T_{\rm s}\Omega\rm\, M_\odot\,{\rm yr}^{-1}$, which could increase to $\sim\,4.0T_{\rm s}\Omega\rm\,M_\odot\,{\rm yr}^{-1}$, corresponding to an energy outflow rate of $2.4T_{\rm s}\Omega\times10^{40}$ -- $1.5T_{\rm s}\Omega\times10^{41}\,\rm erg\,s^{-1}$, where $T_{\rm s}$ is the spin temperature and $\Omega$ is the solid angle of the outflow. Previous optical observations identified an extended emission line region aligned with the radio axis, ionized by the central active galactic nucleus (AGN). Within this region, a warm and ionized outflow with a mass outflow rate of $\sim\,50\rm\, M_\odot\,{\rm yr}^{-1}$ and an energy outflow rate of $\sim1.7\times10^{43}\,\rm erg\,s^{-1}$ was detected. We propose that both the extended emission line region and the optical outflow are results of a synergistic effect between the jet and AGN radiation. The AGN likely exerts negative feedback on the host galaxy, as is evidenced by the gas expulsion by the jet and the high-velocity dispersion of ionized gas observed optically. So far, detections of jet-driven neutral hydrogen outflows remain rare. The high redshift, large outflow radii, substantial mass outflow rate, and energy outflow rate of the neutral hydrogen outflow in 0731+438 expand the known parameter space of such outflows. Finally, we find tentative correlations between the neutral hydrogen mass outflow rate (or energy outflow rate) and the rest-frame \mbox{1.4 GHz} radio power of jets for the known broad neutral hydrogen outflows.

}
   \titlerunning{Jet-gas interaction in 0731+438}
   \authorrunning{Renzhi Su et al.}
   \maketitle
%
\section{Introduction}
\mbox{H{\sc i}}, as the most abundant element in the Universe, plays a fundamental role in galaxy evolution by supplying fuel for star formation. Extensive research has demonstrated its critical involvement in black hole accretion and feedback \citep[e.g.][]{morganti2018}. \mbox{H{\sc i}} absorption, whereby the foreground neutral hydrogen absorbs the background radio emission at a rest frame wavelength of $\sim$ 21 cm, serves as a powerful diagnostic tool for investigating neutral gas distribution in and around galaxies \citep[e.g.][]{allison2012,maccagni2017,aditya2018,gupta2021,allison2022,su2022,chandola2024,hu2025,yoon2025}. Usually, symmetric \mbox{H{\sc i}} absorption profiles centered on the systemic velocity of host galaxies\footnote{In this paper, we define a host galaxy as either the stellar distribution observed optically or the spatial region predominantly occupied by star-forming activity.} regularly trace a rotating disk, while asymmetric and broad \mbox{H{\sc i}} absorption lines reveal outflows driven by jets \citep[e.g.][]{morganti2005,morganti2018}. Studies have demonstrated that observing \mbox{H{\sc i}} absorption in bright radio galaxies is an effective method of understanding the gas kinematics and dynamics in jet-gas interaction and further the feedback by jets \citep[e.g.][]{mahony2013,morganti2013,su2023a,su2023b}. 

Jets can exert powerful feedback on host galaxies \citep[e.g.][]{wagner2011,wagner2012,mukherjee2018}. As jets develop, they would interact with surrounding interstellar gas, transferring substantial energy that modifies gas kinematics and structure, ultimately regulating star formation activity. A famous example is 4C\,12.50 in which, through \mbox{H{\sc i}} absorption observations, a cloud of neutral hydrogen gas was found to be expelled outward by the radio jet, possibly inducing negative feedback \citep{morganti2013}. More importantly, jets can restart for many times throughout the lifetime of a galaxy, implying that jet-driven feedback can repeatedly occur, and hence have a profound impact on the host galaxies, altering their evolution trajectory. However, observations and analysis of jet-gas interaction are not sufficient \citep{mukherjee2025}; an adequate picture of jet-gas interaction and its feedback often involves multiwavelength observations of multiphase gas. To date, the detections of jet-driven neutral hydrogen outflows are scarce, mainly are at low redshift and at parsec or kiloparsec scales \citep{morganti2018,mukherjee2025}, limiting our knowledge.

Here, we report a particular case of jet-driven outflow at $z$ = 2.429. 0731+438 is a type 2 quasar \citep{derry2003}, exhibiting \mbox{FR II} radio morphology with two lobes and a core \citep{carilli1997,morabito2016,sweijen2022,cordun2023}. The radio emission extends to 9.9 arcsec, corresponding to a projected linear distance of 82 kpc \citep{sweijen2022}. Subaru observations reveal biconical lobes of $\rm H\alpha$+[N {\sc ii}] emission lines with an extent of 40 kpc that aligned with the radio axis \citep{motohara2000}, supporting the existence of jet-gas interaction. Previous Westerbork Synthesis Radio Telescope (WSRT) observations indicate a possible faint \mbox{H{\sc i}} absorption line \citep{rottgering1999}. However, the spectrum was severely affected by radio frequency interference (RFI) and was not presented. We therefore proposed upgraded Giant Metrewave Radio Telescope (uGMRT) observations to: 1) test whether the \mbox{H{\sc i}} absorption is real; 2) study the neutral hydrogen gas in the jet-gas interaction.

We organize the paper as follows. We present the observations and data reduction in Sect. \ref{sec:obser_dr} and the observational results in Sect. \ref{sec:results}. Discussions and a summary are in Sects. \ref{sec:discussions} and \ref{sec:summary}, respectively.

\section{Observation and data reduction} \label{sec:obser_dr}
We conducted the uGMRT observations on January 20, 2023. A bandwidth of 25 MHz centered on 414.3 MHz and divided into 2048 channels was used, which provided a velocity coverage of -9000--9000 $\rm km \,s^{-1}$ and a velocity resolution of $\sim$ 8.8 $\rm km \,s^{-1}$. 3C 147 was used for bandpass and flux calibrator. Our target 0731+438 was observed for three scans of 38 minutes each embraced by scans of a gain calibrator, 0702+445. 

The data were calibrated and imaged using CASA \citep{casa2022}. Before calibrating, we first inspected the raw visibilities and removed bad data from shadowing, the beginning and end of each scan, non-working antennas, and RFI. We then performed the following calibration steps: an initial flux density scaling, bandpass calibration, complex gain calibration, scaling the amplitude of the complex gain calibrator, and applying the solutions. Next we further inspected the data and flagged outliers. The above calibration-flagging process was repeated using the flagging table produced in the previous iteration to refine the data. Absorption was clearly visible in the calibrated visibilities over 413.610 -- 414.831 MHz, which corresponds to $z$=2.424--2.434. 

Finally, we split our target visibilities. To image the continuum, we excluded the line channels and binned the visibilities to a coarse resolution of 244 kHz. We made the continuum image with the task tclean. Two iterations of phase-only self-calibrations and one iteration of amplitude-and-phase self-calibration were applied to improve the image quality. To make the spectral cube, we first applied the self-calibration solutions to the original calibrated high-frequency resolution visibilities, then we fit a second-order polynomial in the frequency ranges of 412.390 -- 413.610 MHz and 414.831 -- 416.052 MHz in the visibility domain to subtract the continuum. We next inspected the spectral baselines and flagged those with inadequate continuum subtraction. Finally, we generated the cube using the channels between 412.390 and 416.052 MHz. Since we flagged more baselines when making the cube, we remade the continuum image with the same baselines, which guarantees the same UV coverage for the continuum image and the spectral cube.

\section{Results}\label{sec:results}

\subsection{Continuum images}
\begin{figure}
        \includegraphics[width=\columnwidth]{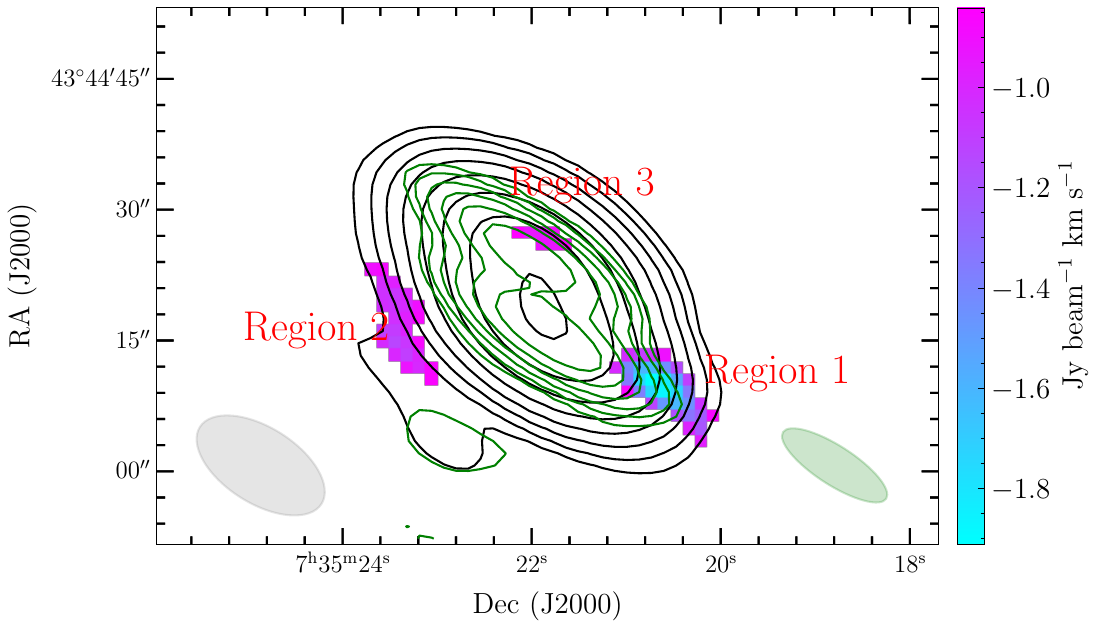}
    \caption{Continuum images and \mbox{H{\sc i}} absorption moment 0 map. The black contours represent the image with natural weighting, which start at $3\sigma$ noise and increase by a factor of 2, while the green contours show the image with uniform weighting, which start at $3\sigma$ noise and increase by a factor of 2. The color image is the \mbox{H{\sc i}} absorption moment 0 map with natural weighting. Only pixels with absorption stronger than 1.5$\sigma$ are shown, where $\sigma$ is the noise of the map. The right color bar shows the depth of the absorption. The map contains three unresolved regions, compared to the beam size. \mbox{H{\sc i}} absorption lines were extracted from the peak absorption pixels of the three regions, which are presented in Fig. \ref{fig:hi_spectra}.      }
    \label{fig:gas_distri}
\end{figure}

Figure \ref{fig:gas_distri} shows the continuum images. The image with natural weighting has a beam size of 16.5 arcsec $\times$ 8.7 arcsec with a position angle (PA) of $57.7^{\circ}$, a peak flux density of 1.67$\pm$0.08 Jy beam$^{-1}$, an integral flux of 2.7$\pm$0.1 Jy, and a root mean square (rms) noise of 4.0 mJy beam$^{-1}$. The image with uniform weighting has a beam size of 14.1 arcsec $\times$ 4.5 arcsec with a PA of $57.2^{\circ}$, a peak flux density of 1.27$\pm$0.06 Jy beam$^{-1}$, an integral flux of 2.6$\pm$0.1 Jy, and a rms noise of 5.0 mJy beam$^{-1}$. Note that the associated uncertainties were derived by assuming the flux calibrator has a flux uncertainty at the 5\% level. Two components are clearly seen in the uniform weighting image, corresponding to the two radio lobes disclosed in other higher-spatial-resolution images \citep{carilli1997,morabito2016,sweijen2022,cordun2023}. The southern lobe is about \mbox{47 kpc} from the core, while the northern lobe is about \mbox{36 kpc} from the core \citep{sweijen2022}.

\subsection{Spectral results}
\begin{figure}
        \includegraphics[width=\columnwidth]{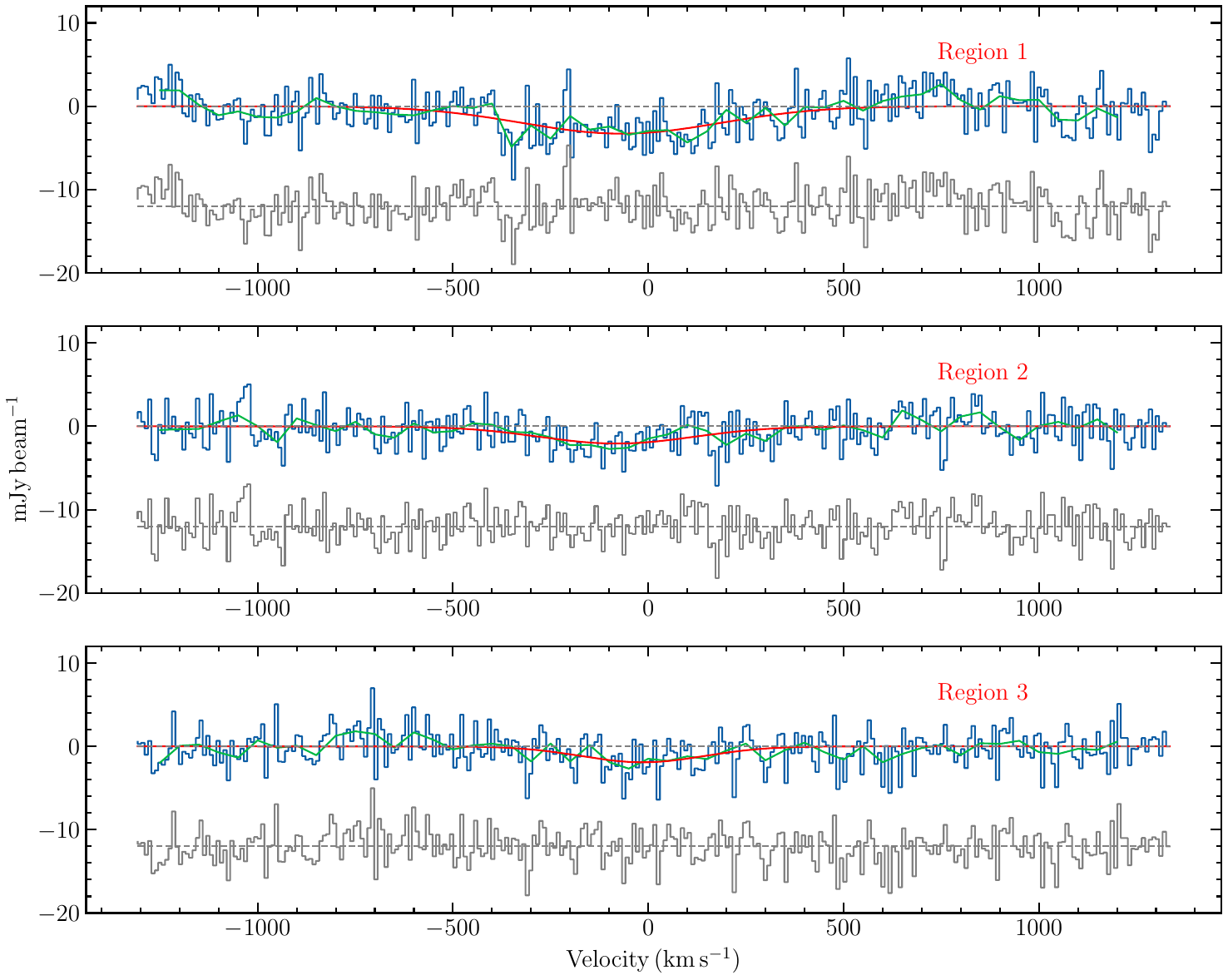}
    \caption{\mbox{H{\sc i}} absorption lines extracted from the peak absorption pixels of the three regions in Fig. \ref{fig:gas_distri}.The blue lines are observed spectra, the green lines are observed spectra resampled to a velocity resolution of 50 $\rm km\,s^{-1}$ using SpecRes \citep{carnall2017}, the red lines are fit Gaussian components, and the gray lines are residuals. }
    \label{fig:hi_spectra}
\end{figure}

We generated the spectral cube with natural weighting, which has an rms noise of 2.16 mJy beam$^{-1}$ with a velocity resolution of $\sim$ 8.8 km s$^{-1}$. To pinpoint the location of the absorption, we made the moment 0 map; see Fig. \ref{fig:gas_distri}. In the map we see three distinct regions where the absorption exceeds 1.5 $\sigma$, where $\sigma$ is the rms noise of the moment 0 map.  Given the beam size, these regions are not resolved and so we extracted spectra from the peak absorption pixels of the regions. The spectra are presented in Fig. \ref{fig:hi_spectra}. 

All spectra are well fit by a single Gaussian function. The fit parameters are listed in Table \ref{tab:spectra_fitting}. The entire line 1 has a significance of $\sim 10\sigma$, whereas lines 2 and 3 have lower significances of $\sim 5\sigma$ and $\sim 4\sigma$ throughout their entirety. For the sake of conservation, we take lines 2 and 3 to be possible detections. Furthermore, the \mbox{H{\sc i}} column density calculated from the line 3 is much lower than those of line 1 and line 2, indicating negligible influence, we so exclude the line 3 in the analysis. However, line 2 exhibits the highest column density, which is retained as a limit in the analysis.

\begin{table*}
\caption{Line parameters. }
\label{tab:spectra_fitting}

\begin{tabular}{lccccccc}
\hline 
Line & $F_{\rm abs,p}$ & $v$ & FWHM & $F_{\rm cont}$ & $\tau_{\rm p}$ & $\tau_{\rm int}$  & $N_{HI}$\\

 &$\rm mJy\,beam^{-1}$ & km\,s$^{-1}$& km\,s$^{-1}$ & $\rm mJy\,beam^{-1}$  & & km\,s$^{-1}$ &atoms\,cm$^{-2}$ \\
\hline \noalign {\smallskip}

Line 1&-3.3$\pm$0.4 & -69.0$\pm$34.6& 599.2$\pm$81.4& 250.6 & 0.013$\pm$0.002  & 8.32$\pm$1.73 & $(1.5\pm0.3)\times10^{19}T_{\rm s}$ \\
Line 2&-2.1$\pm$0.4 &-80.7$\pm$43.5 &467.3$\pm$102.3 & 14.8 & 0.14$\pm$0.03  & 69.70$\pm$21.73 &$(1.3\pm0.4)\times10^{20}T_{\rm s}$  \\
Line 3&-1.9$\pm$0.4 &-14.3$\pm$42.3 &357.6$\pm$100.0 & 878.6 & 0.0022$\pm$0.0005  & 0.87$\pm$0.31 & $(1.6\pm0.6)\times10^{18}T_{\rm s}$ \\

\hline 
\end{tabular}
\tablefoot{Column 1 is the line name; Column 2--4 are fit Gaussian parameters including the peak absorption amplitude, central velocity, and velocity FWHM; Column 5 is the continuum flux density of the pixels from which the lines are extracted; Column 6 is the peak optical depth; Column 7 is the integrated optical depth; Column 8 is the \mbox{H{\sc i}} column density. }
\end{table*}

\section{Discussion} \label{sec:discussions}
\subsection{The detection of \mbox{H{\sc i}} absorption at high redshift}
It is well known that the \mbox{H{\sc i}} absorption detection rate is lower at higher redshift \citep[e.g.][]{curran2008,aditya2018b,su2022,curran2024}. Despite extensive searches \citep[e.g.][]{curran2008,curran2016,curran2017,grasha2019,gupta2021b}, the number of associated \mbox{H{\sc i}} absorption detections at $z>2$ is just five, which are B2 0902+345 at $z=3.397$ \citep{uson1991}, MG J0414+0534 at $z=2.637$ \citep{moore1999}, 8C 0604+728 at $z=3.523$ \citep{aditya2021}, M1540-1453 at $z=2.114$ \citep{gupta2021b,gupta2022}, and NVSS J164725+375218 at $z=2.327$ \citep{chime2025}. Our detection toward 0731+438 at $z=2.429$ is the sixth at $z>2$.

The low detection was attributed to the reliance upon on optical redshift biasing toward the most UV luminous objects at high redshift \citep{curran2008}. \cite{curran2012} subsequently showed that the critical luminosity above which \mbox{H{\sc i}} was not detected, $L_{1216\text{\AA}}>10^{23}\,\rm W\, Hz^{-1}$ or an ionizing photon rate of $Q\sim10^{56}\rm\,s^{-1}$ \citep{curran2017}, was just sufficient to ionize all the neutral gas in a large spiral galaxy. Most recently, \cite{curran2024} has shown an anticorrelation between the \mbox{H{\sc i}} absorption strength and the ionizing photon rate. Using Keck spectroscopic observations, \cite{vernet2001} measured a monochromatic luminosity of $6.3\times10^{41}\,\rm erg\,s^{-1}\,\text{\AA}^{-1} $ at rest frame wavelength $\lambda_{\rm rest} =1500\text{\AA}$ and a continuum slope of $\beta=-1.43$ ($F_{\lambda}\propto \lambda^{\beta}$) between 1500 $\text{\AA}$ and 2000 $\text{\AA}$. Extrapolating the slope to 1216 $\text{\AA}$, we obtained a UV luminosity of $L_{1216\text{\AA}}=4.2\times10^{22}\,\rm W\, Hz^{-1}$, which is below the critical value of $10^{23}\,\rm W\, Hz^{-1}$. This seems to explain why we can detect \mbox{H{\sc i}} absorption in 0731+438. However, as the observed UV luminosity has not been corrected for absorption from the host galaxy, the detailed situation is complicated. 

Subaru narrow-band observations of $\rm H\alpha$+[N{\sc ii}] disclosed biconical lobes with an extent of 40 kpc aligned with the radio axis \citep{motohara2000}. Analysis suggested that the $\rm H\alpha$ emission come from nebula ionized by the strong isotropic UV radiation of the central active galactic nucleus (AGN). The ionizing photon rate required to produce the observed $\rm H\alpha$ luminosity is at least $1.4\times10^{57}\rm\,s^{-1}$, which is much larger than the critical value of $\sim10^{56}\rm\,s^{-1}$ \citep{curran2017}. It seems that the observation is in conflict with theory. 

It is worth noting that the radio source displays an FR II morphology, with a linear size of 82 kpc \citep{sweijen2022}, and that the \mbox{H{\sc i}} absorption takes place outside the host galaxy (see Sect. \ref{sec:gas_role}). While an ionizing photon rate of $Q\sim10^{56}\rm\,s^{-1}$ might be sufficient to ionize the neutral gas within the host galaxy, it would be insufficient to ionize the neutral gas in the halo or circumgalactic medium (CGM).

The detection of \mbox{H{\sc i}} absorption in 0731+438 may inform us on the future observational strategy for \mbox{H{\sc i}} absorption at high redshift ($z>2$). Previous searches for \mbox{H{\sc i}} absorption are biased to compact radio sources, which are still confined in host galaxies, and hence exhibit a higher detection rate compared to extended radio sources \citep[e.g.][]{curran2013,maccagni2017}. While it is a good criterion to search for \mbox{H{\sc i}} absorption at low redshift, it may not be ideal for sources at high redshift due to the preselection of sources with a high rest-frame UV luminosity that could ionize the neutral gas in host galaxies \citep[e.g.][]{curran2008,aditya2016,curran2017,aditya2018,grasha2019}. We suggest that selecting extended radio sources, such as FR I and \mbox{FR II} types, rather than compact radio sources for searches of high-redshift-associated \mbox{H{\sc i}} absorption might be more efficient, aiming to probe the neutral gas in halo or CGM rather than in host galaxies. More observations of extended radio sources at high redshift will test this hypothesis.

\subsection{The role of the neutral gas}\label{sec:gas_role}

Usually, symmetric \mbox{H{\sc i}} absorption profiles centered on the systemic velocity of host galaxies trace regularly rotating disks, while asymmetric and broad \mbox{H{\sc i}} absorption lines reveal outflows driven by jets \citep[e.g.][]{morganti2005,morganti2018}. The gas detected in 0731+438 has two outstanding properties that can help us to know what role they are playing. The first is the location of the gas. As is seen in Fig. \ref{fig:gas_distri}, lines 1 and 2 are associated with the southern radio lobe and at its edge, while line 3 is associated with the northern radio lobe. Both lobes have extended out of the host galaxy, indicating that the detected gas is not from any rotating disk in the host galaxy. The second is the large velocity width. Line 1 has a large velocity span with a full width at half maximum (FWHM) of $\sim\,600\,\rm km\,s^{-1}$. This resembles the kinematics of the outflowing gas driven by the jet that has been detected in, for example, 4C 12.50, 3C 293, and IRAS 10565+2448 \citep{morganti2013,mahony2013,su2023a}. These two properties unambiguously point out that the lines we observed are tracing jet-driven outflows. 

To obtain the \mbox{H{\sc i}} mass outflow rate, we followed the equation \citep{heckman2002} 
\begin{equation}\label{equ:HI_mass_outflow_rate}
\dot{M} = 30 \frac{\Omega}{4\pi} \frac{r_*}{1\,\mathrm{kpc}} \frac{N_{\mathrm{HI}}}{10^{21}\mathrm{cm}^{-2}} \frac{v}{300\,\mathrm{km\,s}^{-1}} \rm M_\odot\,{\rm yr}^{-1}
,\end{equation}

\noindent where the $\Omega$ is the solid angle of outflow, $r_*$ is the outflow radius, $N_{\mathrm{HI}}$ is the \mbox{H\,{\sc i}} column density, and $v$ is the outflow velocity. 

\begin{figure*}
        \includegraphics[width=\columnwidth]{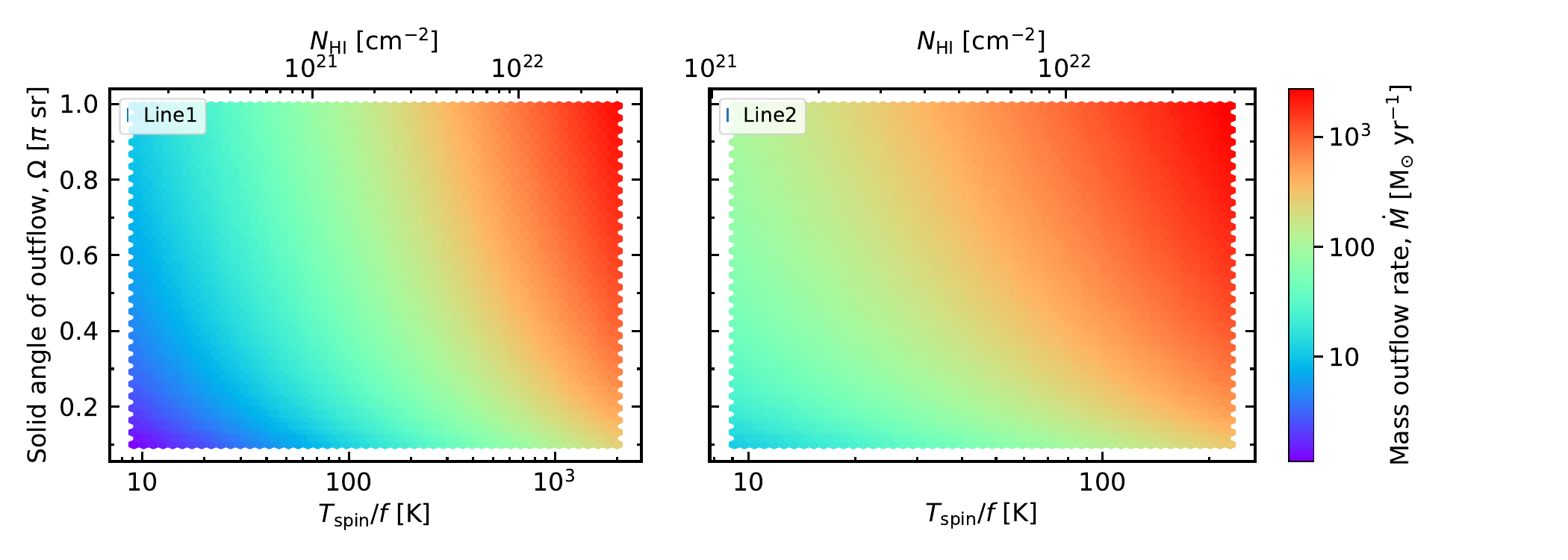}\includegraphics[width=\columnwidth]{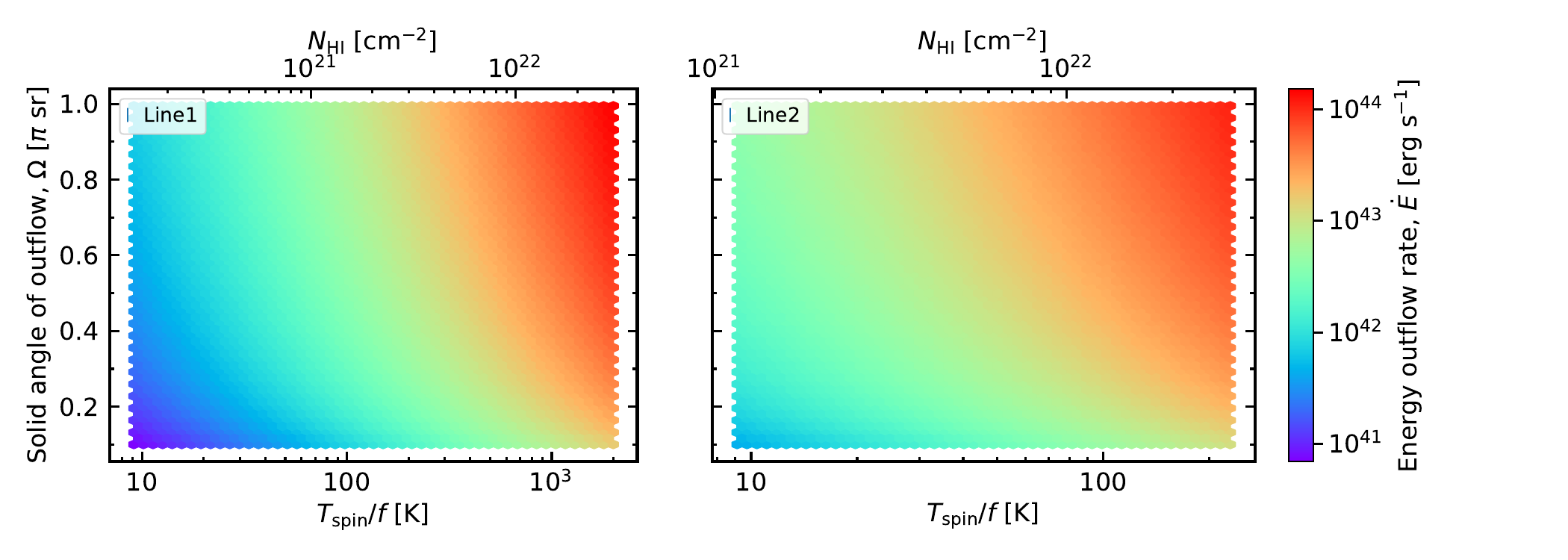}
    \caption{Distributions of \mbox{H{\sc i}} column density, mass outflow rate, and energy outflow rate as a function of $\Omega$ and $T_{\rm spin}/f$ for lines 1 and 2, where $\Omega$ is the solid angle of the outflow and $T_{\rm spin}/f$ is the spin temperature divided by the covering factor. Note that the column density limits are constrained by $T_{\rm spin}/f>T_{\rm cmb}$ at z=2.4 and $\rm N_{HI}<3\times10^{22} $\citep{schaye2001,curran2024}, where $T_{\rm cmb}$ is the temperature of the cosmic microwave background.}
    \label{fig:outflow_m_e}
\end{figure*}

Lines 1 and 2 are associated with the southern jet. Based on the fit line parameters and the continuum flux density of the pixels from where we extracted the lines, we calculated the peak and integrated optical depth, which are listed in Table \ref{tab:spectra_fitting}. The \mbox{H{\sc i}} column density is dependent on spin temperature, $T_{\rm s}$. Line 1 suggests a column density of $(1.5\pm0.3)\times10^{19}T_{\rm s}\,\rm cm^{-2}$, while line 2 suggests a column density one order of magnitude higher. As a compromise, when we estimated the mass outflow rate, we used the column density calculated from line 1 as a lower limit and the column density calculated from line 2 as an upper limit. The southern jet is at a distance of about 47 kpc from the core, which is the outflow radii. The velocity, $v$, was determined from the line fit. Therefore, $\dot{M}=0.387T_{\rm s}\Omega$ $\rm M_\odot\,{\rm yr}^{-1}$ (using line 1), which could increase to $3.925T_{\rm s}\Omega$ $\rm M_\odot\,{\rm yr}^{-1}$ (using line 2). We show the distribution of $\dot{M}$ as a function of $T_{\rm s}$ and $\Omega$ in Fig. \ref{fig:outflow_m_e}. Assuming that $T_{\rm s}$ = 100 K and $\Omega=\pi/2$ for the outflow, the mass outflow rate spans between $\sim\,60.8\rm\, M_\odot\,{\rm yr}^{-1}$ and $\sim\,616.3\rm\,M_\odot\,{\rm yr}^{-1}$.

To understand the impact of the outflow, it is crucial to know the kinetic power of the outflow. We used
\begin{equation}\label{equ:energy_loss_rate}
\dot{E} = 6.34 \times10^{35} \frac{\dot{M}}{2} \left(v^2 + \frac{{\rm FWHM}^2}{1.85}\right) {\rm erg}\,{\rm s}^{-1}, 
\end{equation}
\noindent which accounts for the turbulence \citep{holt2006}, giving $\dot{E}=2.4T_{\rm s}\Omega\times10^{40}$ -- $1.5T_{\rm s}\Omega\times10^{41}$ $\rm erg\,s^{-1}$. We show the distribution of $\dot{M}$ as a function of $T_{\rm s}$ and $\Omega$ in Fig. \ref{fig:outflow_m_e}. Analogously, assuming $T_{\rm s}$ = 100 K and $\Omega=\pi/2$, the $\dot{E}$ is at somewhere between $3.8\times10^{42}$ and $2.4\times10^{43}\,\rm erg\,s^{-1}$.

\subsection{The jet-gas interaction at other wavelengths in 0731+438}
Multiwavelength observations can yield information from multiphase gas. While radio observations provide the information about cold neutral hydrogen gas, optical observations can inform us about the warm ionized gas. The Subaru observations found biconical lobes with an extent of 40 kpc aligned with the radio axis, supporting the existence of jet-gas interaction. Furthermore, Keck spectroscopic observations detected a longer extent of $\rm Ly\,\alpha$ emission along with the radio axis \citep{martin2003}. What is intriguing is that the $\rm Ly\,\alpha$ emission line exhibits two kinematic components, including one narrow component and one broad component (FWHM $>1000\rm\, km\,s^{-1}$). The narrow component has a low speed ($\leq100\rm\, km\,s^{-1}$) and is distributed from the northern halo to the southern halo, whereas the broad component has a higher velocity and is only detected in the northern halo. \cite{martin2003} considered the narrow component as part of quiescent halo gas. The distinct kinematics of the broad component suggests that it is an outflow driven by the jet. 

The gas environments of the northern and southern halos differ. The $\rm H\alpha$+[N{\sc ii}] lines and $\rm Ly\,\alpha$ line are more extended and disturbed to the north, suggesting a more diffuse environment. This is also supported by the steeper spectral index of the northern radio lobe compared to that of the southern radio lobe \citep{morabito2016,sweijen2022}. \cite{motohara2000} measured an electron density of 38 $\rm cm^{-3}$ for the northern gas and 68 $\rm cm^{-3}$ for the southern gas. 

We estimated the mass outflow rate and energy outflow rate of the ionized gas using the broad $\rm Ly\,\alpha$ component \citep{martin2003}. The speed and luminosity of the broad $\rm Ly\,\alpha$ component were not given in \cite{martin2003} and so we estimated these based on the known luminosity, $1.2\times10^{44}\rm\,erg\,s^{-1}$, of the narrow $\rm Ly\,\alpha$ component and the $\rm Ly\,\alpha$ spectra presented in Fig. 5 of the paper. By inspecting the $\rm Ly\,\alpha$ spectra, we found that the broad component has a similar luminosity as the narrow component and an approximate speed of $700\rm\, km\,s^{-1}$. We then followed \cite{holt2006}:
\begin{equation}\label{equ:lya_mass_outflow_rate}
\dot{M}_{\rm ionized} = \frac{3L({\rm Ly\,\alpha})m_{\rm p}v_{\rm out}}{N\alpha_{\rm Ly\,\alpha}^{\rm eff}h\nu({\rm Ly\,\alpha})r}, 
\end{equation}
\noindent where $L({\rm Ly\,\alpha})$ is the luminosity of $\rm Ly\,\alpha$ line, $m_{\rm p}$ is the mass of proton, $v_{\rm out}$ is the outflow speed, $N$ is the electron density, $\alpha_{\rm Ly\,\alpha}^{\rm eff}\sim1.7\times10^{-13}\rm\,cm^{3}\,s^{-1}$ is the effective $\rm Ly\,\alpha$ coefficient \citep{osterbrock1989}, $h$ is the Planck constant, $\nu(\rm Ly\,\alpha)$ is the frequency of $\rm Ly\,\alpha$ line, and $r$ is the outflow radius. By adopting the distance between the northern radio lobe and the central radio core as the outflow radius, we estimate a mass outflow rate of $\sim\,50\rm\, M_\odot\,{\rm yr}^{-1}$ for the ionized outflow, corresponding to an energy outflow rate of $\sim1.7\times10^{43}\,\rm erg\,s^{-1}$ using Eq. \ref{equ:energy_loss_rate}. These are comparable to those of the outflow detected by \mbox{H{\sc i}} absorption. We suggest that future integral field unit (IFU) observations with a high spatial resolution could provide a more comprehensive understanding of the distribution and kinematics of the ionized gas including the outflowing component.

\subsection{Feedback}

Powerful jets are a main source of feedback, which could inject considerable energy into the host galaxy, perturbing gas kinematics, driving outflows, and hence regulating galaxy evolution. In hydrodynamical simulations of jet-gas interactions, induced feedback becomes efficient in dispersing gas when the ratio between jet power and AGN Eddington luminosity exceeds $\eta\equiv P_{\rm jet}/L_{\rm E}\sim 10^{-4}$ \citep{wagner2011,wagner2012}. The observed flux density of 2.7 Jy gives a rest-frame 1.4 GHz radio power of $3.7\times10^{28}\,\rm W\,Hz^{-1}$, from which we estimate a jet power of $P_{\rm jet}=2.8\times10^{47}\,\rm erg\,s^{-1}$ \citep{cavagnolo2010}. To estimate the Eddington luminosity, we also need the mass of the supermassive black hole (SMBH). Assuming $\eta = 10^{-4}$ gives $L_{\rm E} = 2.8\times10^{51}\,\rm erg\,s^{-1}$, corresponding to a SMBH mass of $2.5\times10^{13}\, M_\odot$. This is unreasonably large compared to the expected $10^{7-8}\, M_\odot$, and applying the expected masses gives $\eta \gtrsim 10$, suggesting that the jet is capable of inducing negative feedback.

A prominent observational property is the large-scale warm ionized gas emission aligned with the radio axis, ionized by the central AGN \citep{motohara2000,martin2003}. This is a direct result of feedback. A simple picture that can be established is that a tunnel was created due to the movement of the jet so that the central AGN radiation can directly have an effect on the gas surrounding the tunnel. This is why we see that the ionized gas is aligned with the radio axis, whereas the gas in other regions is still largely immune from AGN radiation.  In the “two-stage” feedback model proposed by \cite{hopkins2010}, the AGN can easily disperse gas at large radii and inhibit star formation. In this model, a weak wind or outflow first breaks up the stability of dense gas in pressure equilibrium by reducing its ambient pressure. Consequently, the gas will expand perpendicularly to the outflow, which will increase the effective cross section of the gas material and make it much more susceptible to AGN radiation. In the case of 0731+438, the passage of the jet opens a putative tunnel, which could also remove the pressure equilibrium of the surrounding gas so that the subsequent AGN radiation can accelerate the gas, giving a large velocity dispersion. The observed $\rm Ly\,\alpha$ line with large velocity FWHM (a narrow component with FWHM of $600\,\rm km\,s^{-1}$ and a broad component with FWHM of $1000\,\rm km\,s^{-1}$) supports this scenario \citep{martin2003}. Therefore, under this condition, the feedback is negative (suppressing star formation). What is interesting is that this provides an example of a synergistic effect between the jet and AGN radiation, showing they can work together to produce feedback for the host galaxy.

It is well known that the CGM is important for galaxy growth by supplying new gas for star formation \citep[e.g.][]{tumlinson2017}. The outflow in 0731+438 is expelling gas away from the host galaxy, which highlights the role of the jet in impeding this process. In short, the AGN feedback (including jet-induced and AGN-radiation-induced feedback) in 0731+438 is most likely negative.

\begin{figure*}
        \includegraphics[width=\textwidth]{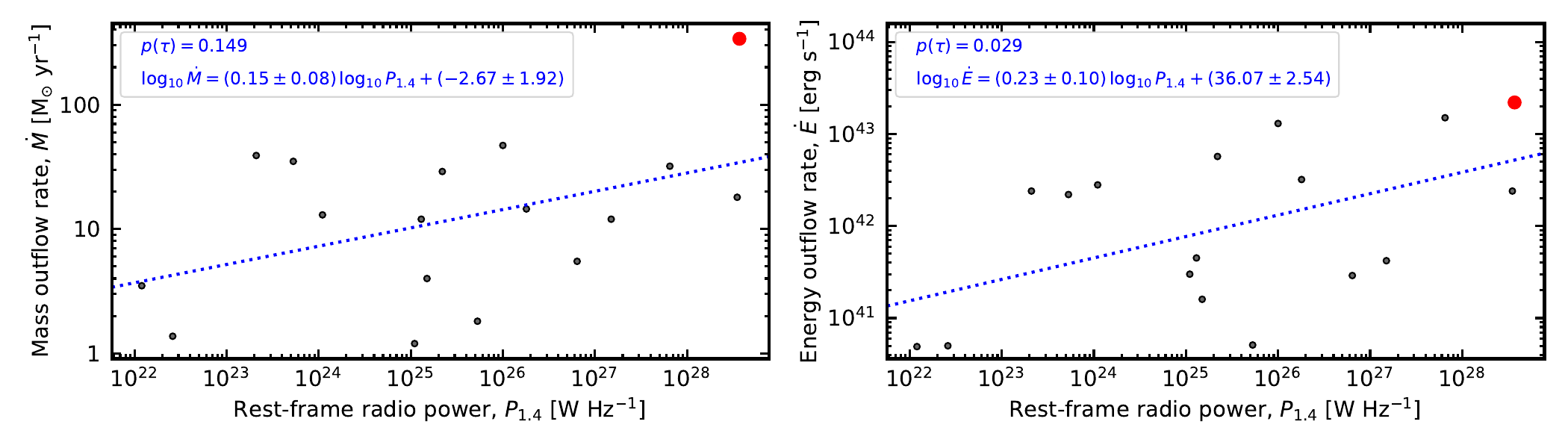}        
    \caption{Correlations between the neutral hydrogen outflow properties, mass outflow rate, and energy outflow rate, and the rest-frame 1.4 GHz radio power of jets. The red markers represent 0731+438. The blue lines are linear fits.}
    \label{fig:pj_rl_oe}
\end{figure*}

\subsection{Comparison with low redshift jet-driven outflows}
The detection of jet-gas interaction is not rare, usually disclosed through the alignments between radio jets and optical extended emission lines \citep{tadhunter2000,martin2003}. However, the detections of jet-driven neutral hydrogen outflows are rare; see Table 1 in \cite{morganti2018}. Since then, further detections have been made in IRAS 10565+2448 \citep{su2023a}, SDSS J145239.38+062738.0 \citep{su2023b}, and 0731+438 in this work. They are mostly seen at $z<1$ and 0731+438 is the highest-redshift example of these.

It is interesting to compare the outflow in 0731+438 with previous known outflows. We used the sources with a known mass outflow rate in Table 1 of \cite{morganti2018}, in addition to IRAS 10565+2448 \citep{su2023a} and SDSS J145239.38+062738.0 \citep{su2023b}. Note that some sources are detected with multiwavelength outflows, whereas others are not. For a relatively fair comparison, we only used the neutral hydrogen outflows detected with \mbox{H{\sc i}} absorption. For the neutral hydrogen mass outflow rate of NGC 1266, we acquired it from \cite{alatalo2011}. Further, for IRAS 10565+2448, we assumed $T_{\rm s} = 100$ K, which gives a neutral hydrogen mass outflow rate of $39\rm\, M_\odot\,{\rm yr}^{-1}$. For 0731+438, we assumed $T_{\rm s} = 100$ K and $\Omega=\pi/2$. Moreover, for sources with a range of reported mass outflow rates, we used the central value of the range. We also calculated the rest frame 1.4 GHz monochromatic luminosity of this sample with using the NASA/IPAC Extragalactic Database (NED).

Finally, their mass outflow rate is plotted as a function of their rest frame 1.4 GHz monochromatic luminosity $P_{1.4}$ in Fig. \ref{fig:pj_rl_oe}. It is clear that 0731+438 has the highest $P_{1.4}$ and also the highest mass outflow rate. In addition, it seems that sources with higher $P_{1.4}$ have a higher mass outflow rate. This is not surprising as in principle more powerful jets should drive more powerful outflows. To test it, we performed a linear fit, which gives the relation
\begin{equation}\label{equ:mor_1.4lumi}
{\rm Log}\,\dot{M}=(0.15\pm0.08){\rm Log}\,P_{1.4} - (2.67\pm1.92)
.\end{equation}This fitting has a Kendall’s tau $p(\tau)$ of 0.149, indicating a possible correlation.

Furthermore, we also compared the energy outflow rate with the $ P_{1.4}$. We calculated the energy outflow rate in a uniform way with Eq. \ref{equ:energy_loss_rate}, except for SDSS J145239.38+062738.0, whose energy outflow rate was directly adopted from \cite{su2023b}. Following \cite{morganti2005}, we used the half of the full width at zero intensity (FWZI/2) of the blueshifted component as the outflow velocity, $v$, and FWZI/2 as the FWHM. For sources with a range of reported energy outflow rates, we used the central value of the range. The energy outflow rate as a function of the $P_{1.4}$ is plotted in Fig. \ref{fig:pj_rl_oe}. The linear fit gives the relation
\begin{equation}\label{equ:eor_1.4lumi}
{\rm Log}\,\dot{E}=(0.23\pm0.10){\rm Log}\,P_{1.4} + (36.07\pm2.54)
.\end{equation}This fitting has a Kendall’s tau $p(\tau)$ of 0.029, indicating a tentative correlation.

Note that this sample is far from complete and may be affected by biases related to host galaxy properties. A larger and more representative sample is needed to robustly test these correlations, which should currently be considered as suggestive rather than conclusive.

\section{Summary} \label{sec:summary}

We observed 0731+438 at $z=2.429$ with uGMRT observations and obtained the below results.

\begin{enumerate}
    \item The uniform-weighting continuum image exhibits two components corresponding to two lobes in the \mbox{FR II} galaxy 0731+438 in previous high-spatial-resolution observations. We detected \mbox{H{\sc i}} absorption toward 0731+438.

    \item  To date, this represents the third-highest redshift at which associated HI absorption has been detected. The central AGN exhibits an ionizing photon rate of at least $1.4\times10^{57}\,\rm s^{-1}$, which is higher than the critical value of $\sim10^{56}\,\rm s^{-1}$ above which the neutral hydrogen in host galaxy could be all ionized, resulting in no \mbox{H{\sc i}} absorption detection. Our \mbox{H{\sc i}} absorption detection in 0731+438 suggests that selecting extended radio sources rather than compact radio sources might be more efficient for detecting \mbox{H{\sc i}} absorption at high redshift. 
    
    \item This is so far the highest-redshift jet-driven neutral hydrogen outflow detected through \mbox{H{\sc i}} absorption. To our knowledge, this outflow has the largest radii ever reported. The outflow has a mass outflow rate of $0.4T_{\rm s}\Omega\rm\, M_\odot\,{\rm yr}^{-1}$ (using line 1), which could go up to $4.0T_{\rm s}\Omega\rm\,M_\odot\,{\rm yr}^{-1}$ (using line 2). These correspond to an energy outflow rate of $2.4T_{\rm s}\Omega\times10^{40}$ $\sim$ $1.5T_{\rm s}\Omega\times10^{41}\,\rm erg\,s^{-1}$.

    Moreover, previous optical observations found a jet-driven, warm, and ionized outflow with a mass outflow rate of $\sim\,50\rm\, M_\odot\,{\rm yr}^{-1}$, corresponding to an energy outflow rate of $\sim1.7\times10^{43}\,\rm erg\,s^{-1}$. Given the outflow properties, they expand the parameter space of jet-driven outflows. 
    
    \item The AGN in 0731+438 should produce a negative feedback. The alignment between the extended optical emission line and the radio axis suggests a synergistic effect between the jet and AGN radiation, showing that they can work together to produce feedback to the host galaxy. In addition, this case highlights the role of the jet in preventing CGM accretion to the host galaxy, impeding this galaxy growth.

    \item Tentative correlations were found between the neutral hydrogen mass outflow rate (or energy outflow rate) and the rest-frame 1.4 GHz radio power of jets. However, a larger and less biased sample is required to further test these correlations.
    \end{enumerate}

\begin{acknowledgements}
We thank the anonymous referee for the comments that improve the paper. We thank the staff of the GMRT that made these observations possible. GMRT is run by the National Centre for Radio Astrophysics of the Tata Institute of Fundamental Research. This research has made use of the NASA/IPAC Extragalactic Database, which is funded by the National Aeronautics and Space Administration and operated by the California Institute of Technology. MFG is supported by the National Science Foundation of China (grant 12473019), the Shanghai Pilot Program for Basic Research-Chinese Academy of Science, Shanghai Branch (JCYJ-SHFY-2021-013), the National SKA Program of China (Grant No. 2022SKA0120102), and the China Manned Space Project with No. CMS-CSST-2025-A07. RZS acknowledges the support from the National SKA Program of China (No. 2025SKA0130100), the Shanghai Super Postdoctoral Incentive Program (No. 2025322), and the China Postdoctoral Science Foundation (Grant No. 2025M783232). M. Glowacki is supported by the Australian Government through UK STFC Grant ST/Y001117/1. M. Glowacki acknowledges support from the Inter-University Institute for Data Intensive Astronomy (IDIA). IDIA is a partnership of the University of Cape Town, the University of Pretoria and the University of the Western Cape. YJC is supported by 2023 Shanghai Oriental Talents Program. 

\end{acknowledgements}

%
%
\bibliographystyle{aa} 
\bibliography{aa57997-25} 

\end{document}